\documentclass[aps,showpacs,twocolumn,floatfix,nofootinbib]{revtex4}
\usepackage{graphicx}
\usepackage{epsfig}

\begin{document}

\title{Scrambling of Hartree--Fock Levels as a universal Brownian--Motion
Process}
\author{Y. Alhassid$^1$, H. A. Weidenm\"uller$^2$ and A. Wobst$^3$}
\affiliation{$^{1}$Center for Theoretical Physics, Sloane Physics Laboratory, Yale
University, New Haven, CT 06520, USA\\
$^2$Max-Planck-Institut f\"ur Kernphysik, D-69029 Heidelberg, Germany\\
$^3$Institut f\"ur Physik, Universit\"at Augsburg, 86135 Augsburg, Germany}

\begin{abstract}
We study scrambling of the Hartree--Fock single--particle levels and wave
functions as electrons are added to an almost isolated diffusive or chaotic
quantum dot with electron--electron interactions. We use the generic
framework of the induced two--body ensembles where the randomness of the
two--body interaction matrix elements is induced by the randomness of the
eigenfunctions of the chaotic or diffusive single--particle Hamiltonian.
After an appropriate scaling of the number of added electrons, the
scrambling of both the HF levels and wave functions is described by a
universal function each. These functions can be derived from a parametric
random matrix process of the Brownian--motion type. An exception to this
universality occurs when an empty level just gets filled, in which case
scrambling is delayed by one electron. An explanation of these results is
given in terms of a generalized Koopmans' approach.
\end{abstract}

\maketitle

\emph{Introduction.} The single--particle levels and eigenfunctions of a
chaotic or diffusive quantum dot without electron--electron interactions
follow the predictions of canonical random--matrix theory (RMT) within a
band of $\sim g$ levels around the Fermi energy, where $g$ is the
dimensionless Thouless conductance~\cite{rmt98,rmp00}. In an almost isolated
dot, the screened Coulomb interaction $V$ between electrons must be taken
into account. In the constant--interaction (CI) model, $V$ is approximated
by the charging energy (a constant for a fixed number of electrons), and a
single--particle picture and RMT can again be used. In this paper, we go
beyond the CI model and use the Hartree--Fock (HF) approximation. This
approximation uses a single--particle picture and yet takes into account
both chaotic or diffusive motion \emph{and} interaction effects beyond the
CI model. The HF equations depend on the one--body density and must be
solved self--consistently. The resulting HF single--particle levels and
eigenfunctions depend on the number of electrons on the dot and, therefore,
change (are \textquotedblleft scrambled\textquotedblright) as electrons are
added to the dot~\cite{Patel98b,ag}. An
experimental signature of scrambling is the saturation of the number of
correlated Coulomb--blockade conductance peaks as~temperature $T$\ increases
(without scrambling, that number should rise linearly with $T$) \cite%
{Patel98b}.

Chaotic or diffusive motion of the electrons induces randomness in the
two--body matrix elements of $V$.
For spinless electrons, scrambling is due to the fluctuating part of these
matrix elements (the average yields the CI model). The fluctuations depend
on $g$ and on the geometry of the dot~and are, therefore, non--universal
\cite{blanter97}. Here we evaluate two statistical measures of 
scrambling and find them to be also non--universal. This is in accord with expectations based on Koopmans'
limit~\cite{koopmans34}, where the HF wave functions are assumed to remain
unchanged as an electron is added to the dot and where the rate of spectral
scrambling depends on $g$~\cite{ag}. In spite of this apparent
non--universality, the mechanism of scrambling does possess universal
features. Indeed, we show that there exists a (non--universal) scaling of
the number of added electrons such that our statistical measures of
scrambling become universal. Scrambling can then be described as a Gaussian
random process (GP) of the Brownian--motion type with the proper symmetry~%
\cite{alhassid95}. An exception occurs when an empty HF level becomes
filled. We support our results analytically in terms of a generalized
Koopmans' approach.

\emph{Approach.} For a fixed number of spinless electrons, the induced
ensembles that characterize chaotic or diffusive systems with interactions
were classified and studied in Ref.~\cite{alhassid05}. Here we address their
properties as electrons are added to the dot. We use a basis of eigenstates
of a single--particle RMT Hamiltonian $h^{(0)}$ of orthogonal symmetry with $%
a_{i}$ and $a_{i}^{\dagger }$\ the associated destruction and creation
operators and $\epsilon _{i}$ the single--particle energies. The latter obey
Wigner--Dyson statistics. The anti--symmetrized two--body matrix elements $%
v_{ij;kl}^{A}$ of $V$ in that same basis are assumed to be
Gaussian--distributed random variables with variances $\sigma ^{2}$
characterized by a parameter $u^{2}$. In a diffusive dot $u^{2}\propto
\Delta ^{2}/g^{2}$ where $\Delta $ is the mean single--particle level
spacing. As an example we take the third orthogonal induced ensemble defined
in Ref.~\cite{alhassid05} for which $\sigma ^{2}(v_{ij;ij}^{A})=4u^{2}$. The
Hamiltonian is
\begin{equation}
H=\sum_{\alpha }\epsilon _{i}a_{i}^{\dagger }a_{i}+{\frac{1}{4}}%
\sum_{ijkl}v_{ij;kl}^{A}a_{i}^{\dagger }a_{j}^{\dagger }a_{l}a_{k}\;.
\label{Hamiltonian}
\end{equation}%
Our numerical results are for $m=40$ single--particle levels and 10,000
realizations. For each realization~(\ref{Hamiltonian}) of the ensemble, we
solve the HF equations for a range of numbers $n$ of electrons and obtain
the single--particle HF levels $\epsilon _{\alpha }^{(n)}$ and wave
functions $\psi _{\alpha }^{(n)}$.

\begin{figure}[h]
\centerline{\includegraphics[scale=0.9]{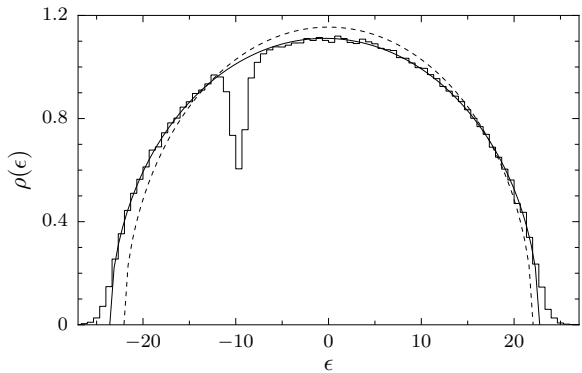}}
\caption{ The average HF level density $\protect\rho (\protect\epsilon )$
for $n=10$ electrons and $u=0.1$ (histogram) is compared with an empirical
fit (see text, solid line) and Wigner's semicircle ($u=0$, dashed line).}
\label{deltarho}
\end{figure}

\emph{Unfolding the HF single--particle levels.} Unfolding is neccessary to
identify unambiguously the local spectral fluctuations. We calculate the
ensemble average of the HF single--particle level density (histogram in Fig.~%
\ref{deltarho}) and fit the result with a function $\rho (\epsilon )$ that
continues smoothly across the HF gap (solid line). We use $\rho (\epsilon
)=(c/3m)\sqrt{12m^{2}-4\pi ^{2}(a+b\epsilon )^{2}}$. For $a=0$, $b=1$ and $%
c=1$ this is just Wigner's semi--circle that corresponds to the choice $%
\sigma ^{2}(h_{\alpha \beta }^{(0)})=3m/4\pi ^{2}$ (dashed line in Fig.~\ref%
{deltarho}). With $\Delta (\epsilon )=1/\rho (\epsilon )$ we define
the unfolded HF single--particle energies as $\tilde{\epsilon}_{\alpha
}=\epsilon _{\alpha }/\Delta (\epsilon _{\alpha })$.

\emph{Scrambling.} We start from a dot with $n=10$ electrons and add $\delta
n$ electrons. For every HF level $\alpha $, we define two statistical
measures for scrambling. Spectral scrambling is measured by the variance of
the change of the unfolded level energy~\cite{ag},
\begin{equation}
\sigma ^{2}(\alpha ,\delta n)=\overline{{\left( \tilde{\epsilon}_{\alpha
}^{(n+\delta n)}-\tilde{\epsilon}_{\alpha }^{(n)}\right) }^{2}}\;,
\label{vle}
\end{equation}
whereas wave--function scrambling is measured by the average of the squared
wave--function overlap
\begin{equation}
o_{\alpha }^{(\delta n)}=\overline{|\langle \psi _{\alpha }^{(n)}|\psi
_{\alpha }^{(n+\delta n)}\rangle |^{2}}\;.  \label{sqwf}
\end{equation}
The bar denotes the ensemble average. Complete spectral scrambling is
achieved when $\sigma (\alpha ,\delta n)\sim 1$, i.e., when $\sigma $ is
comparable to the mean level spacing of the unfolded spectrum.

\begin{figure}[h]
\centerline{\includegraphics[scale=0.9]{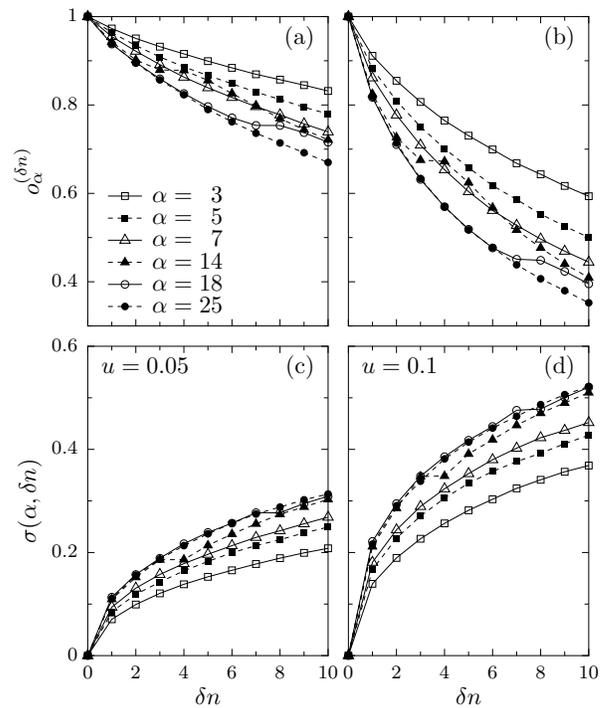}}
\caption{Top panels: Squared wave--function overlap $o_\alpha^{(\delta n)}$
 versus number $\protect\delta n$ of added electrons for several HF levels
$\protect\alpha $. Bottom panels: Standard deviation $\sigma(\alpha ,\delta n)$ of the change in the HF energy of level $
\alpha $ versus $\delta n$. \textquotedblleft Steps" are
observed across the HF gap. The left (right) panels correspond to an
interaction parameter $u=0.05$ ($u=0.1$). In all cases the starting number
of electrons is $n=10$.}
\label{unscaled}
\end{figure}

As $|u|$ increases, the fluctuations of the $v_{ij;kl}^{A}$'s increase, too,
the HF levels and wave functions should, thus, vary more strongly with $%
\delta n,$ and scrambling should increase. And indeed, $o_{\alpha }^{(\delta
n)}$ (top panels of Fig.~\ref{unscaled}) decreases monotonically and $\sigma
(\alpha ,\delta n)$ (bottom panels) increases monotonically with $|u|$.
In Fig.~\ref{unscaled} the levels $\alpha =3,5,7$ ($\alpha =25$) are filled
(empty) in the initial dot and remain so for all values of $\delta n\leq 10$
shown whereas levels $\alpha =14$ and $\alpha =18$ are initially empty and
become filled for $\delta n=4$ and $\delta n=8$, respectively. Here the
scrambling behavior differs qualitatively from that elsewhere: both $\sigma
(\alpha ,\delta n)$ and $o_{\alpha }^{(\delta n)}$ remain approximately
unchanged when the empty level $\alpha $ becomes filled, displaying
 a flat section or a \textquotedblleft
step\textquotedblright . At other values of $\delta n$, both scrambling
measures display their usual monotonic behavior. It is obvious from Fig.~\ref%
{unscaled} that scrambling depends on the particular level and on $u$ and
is, thus, not universal.

\begin{figure}[t!]
\centerline{\includegraphics[scale=0.9]{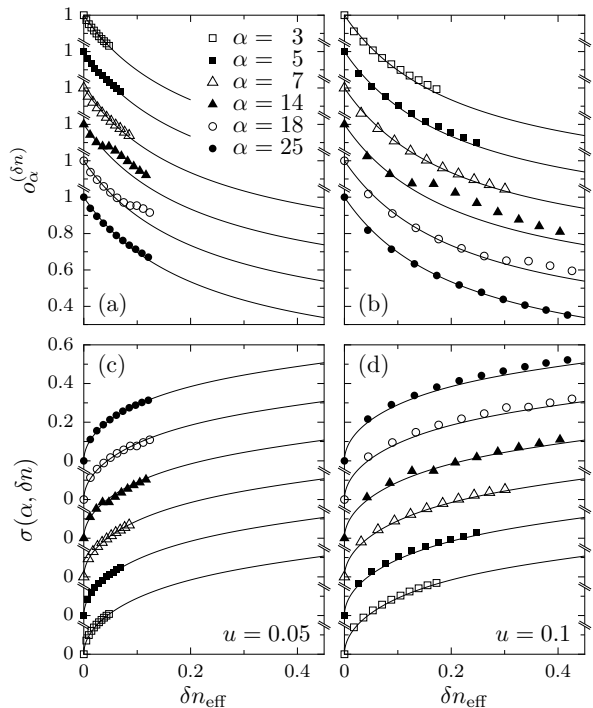}}
\caption{Same data as in Fig.~\ref{unscaled} but as a function of $%
\delta n_{\mathrm{eff}}$ (Eq.~(\ref{deltan})). The solid
lines are predicted from a continuous Gaussian process with $\eta =
1 $. To avoid overlap, the curves are shifted with respect to each other.
The agreement is quite remarkable except when an empty level becomes filled
(i.e., for $\alpha = 14, 18$).}
\label{scaled}
\end{figure}

\emph{Universal scrambling. }For fixed $n$,\ we have shown~\cite{alhassid05}
that the HF ensemble obeys in part universal random--matrix statistics. We
now show that upon proper rescaling of $\delta n$,\ the scrambling measures
also display universal behavior. For reasons given below, we define the
scaled parameter
\begin{equation}
\delta n_{\mathrm{eff}}\approx \frac{4u^{2}}{\Delta ^{2}(\epsilon _{\alpha })%
}\delta n  \label{deltan}
\end{equation}%
and replot in Fig.~\ref{scaled} the data of Fig.~\ref{unscaled} versus $%
\delta n_{\mathrm{eff}}$.
The solid lines represent the same two universal functions, one for the wave
function overlap, and the other for the level variance. These are defined
below.
In contrast to Fig.~\ref{unscaled}, all curves for the same scrambling
measure now show impressive agreement with the universal one. Universality
is violated only for those values of $\delta n$ where a level changes from
empty to filled (steps in the scrambling measures for the levels $\alpha
=14,18$).

\emph{Gaussian random processes (GP).} Our scaling equation~(\ref{deltan})
and the universal correlators can be derived from GP. Continuous GP \cite%
{alhassid95} depend upon an external parameter $x$\ and are known to
describe non--interacting chaotic or diffusive systems which, upon rescaling
$x$, display universal parametric level correlations~\cite{simons93}. For
small values of $|x-x^{\prime }|$, the random Hamiltonian $h_{jk}(x)$ with
orthogonal symmetry obeys
\begin{equation}
\overline{h_{ij}(x)h_{kl}(x^{\prime })}\propto (\delta _{ik}\delta
_{jl}+\delta _{il}\delta _{jk})[1-\kappa |x-x^{\prime }|^{\eta }+\ldots ]%
\text{ }.  \label{GP}
\end{equation}%
This defines the continuous GP with exponent $\eta $~\cite{attias95}. 
The usual parametric
random--matrix models discussed in the literature are \emph{differentiable}
GP for which $\eta =2$. In contradistinction, we will show that in the
 HF approximation, the addition of electrons corresponds to
a GP with $\eta =1$. This exponent is characteristic of Brownian motion
(where time plays the role of the parameter $x$).

\emph{Generalized Koopmans' approach.} To understand the universality of
scrambling we use a generalized Koopmans' approach. With $\alpha $ denoting
the single-particle eigenstates of the $n$--electron HF Hamiltonian,
that Hamiltonian is given by
\begin{equation}
h_{\alpha \gamma }^{(n)}=h_{\alpha \gamma }^{(0)}+\sum_{\beta
=1}^{n}v_{\alpha \beta ;\gamma \beta }^{A}=\delta _{\alpha \gamma }\epsilon
_{\alpha }^{(n)}  \label{HF-n}
\end{equation}%
where the sum is over the lowest $n$ levels. In the same basis, the HF
Hamiltonian for $(n+1)$ electrons is
\begin{equation}
h_{\alpha \gamma }^{(n+1)}=h_{\alpha \gamma }^{(0)}+\sum_{\beta \delta
}v_{\alpha \beta ;\gamma \delta }^{A}\rho _{\delta \beta }\;,
\end{equation}%
where $\rho _{\delta \beta }$ is the single--particle density matrix (a
projector onto the
lowest $(n+1)$ eigenstates of the HF problem with $(n+1)$ electrons). We
expand $h_{\alpha \gamma }^{(n+1)}$ around $h_{\alpha \gamma }^{(n)}$ up to first order in  $v$.
Using $\rho _{\delta \beta }=\delta _{\delta \beta }\Theta (n+1-\beta )+O(v)$
and Eq.~(\ref{HF-n}) we find
\begin{equation}
h_{\alpha \gamma }^{(n+1)}\approx h_{\alpha \gamma }^{(n)}+w_{\alpha \gamma
}^{(n)}\;\; \mathrm{where}\;\; w_{\alpha \gamma }^{(n)}\equiv v_{\alpha n+1;\gamma
n+1}^{A}\ .  \label{HFK}
\end{equation}%
The approximation leading to Eq.~(\ref{HFK}) is tantamount to ignoring the
change in the HF wave functions that occurs when an electron is added to the
dot. When restricted to the diagonal elements of Eq.~(\ref{HFK}),
this neglect is known as Koopmans' approximation. Hence, our Eq.~(\ref{HFK})
can be considered a generalized Koopmans' relation.

\emph{Brownian Motion.} To determine the statistical properties of the
matrix $w$, we use the last Eq.~(\ref{HFK}). For small $|u|$, the
interaction matrix elements in the HF basis were shown~\cite{alhassid05} to
be uncorrelated Gaussian random variables with variances that are very close
to their input values in the basis used in Eq.~(\ref{Hamiltonian}). This
implies that the $w_{\alpha \gamma }$'s are uncorrelated Gaussian random
variables with mean values zero and variances $\sigma ^{2}(w_{\alpha \gamma
}^{(n)})=2(1+\delta _{\alpha \gamma })(1-\delta _{\alpha n+1})(1-\delta
_{n+1\gamma })u^{2}$. The Kronecker symbols in the last two brackets are due
to antisymmetrization and imply that the matrix $w$ carries zeros in the row
and column labelled $(n+1)$. Except for this fact the matrices $w^{(n)}$
form a Gaussian orthogonal ensemble. The single--particle HF wave functions
(which determine $w^{(n)}$) are not correlated with the eigenvalues $%
\epsilon _{\alpha }^{(n)}$. Therefore, the matrices $w^{(n)}$ and $h^{(n)}$
are uncorrelated. Thus, for small $|u|$ and except for effects of the HF
gap, the set of HF Hamiltonians $h^{(n)},h^{(n+1)},\ldots $ with a discrete
parameter $n$ corresponds approximately to a discrete orthogonal GP
(see Ref.~\cite{attias95}). We use the fact that such a GP can be embedded
into a continuous orthogonal GP $h(x)$ by the requirement that $%
h^{(n)}=h(x_{n})$ for a set of discrete points $x_{n}$ with equal spacings $%
\delta x=x_{n+1}-x_{n}$.

The exponent $\eta $ of that GP is determined by the fact that the matrices $
w^{(n)}$ and $w^{(n+1)}$ are approximately uncorrelated. When we use
Eq.~(\ref{HFK}) for both matrices and approximate the HF basis for $(n+1)$
electrons by that for $n$ electrons, this follows from the fact that the
matrix elements $v_{\alpha n+2;\alpha n+2}^{A}$ and $v_{\alpha n+1;\alpha
n+1}^{A}$ are uncorrelated in that basis. We have checked numerically that $%
w^{(n)}$ and $w^{(n+1)}$ remain essentially uncorrelated even when the
matrix elements $v_{\alpha n+2;\alpha n+2}^{A}$ are evaluated in the HF
basis of $(n+1)$ electrons. We conclude that our discrete GP is
characterized by $\overline{[h(x_{n+2})-h(x_{n+1})][h(x_{n+1})-h(x_{n})]}%
\approx \overline{w^{(n+1)}w^{(n)}}\approx 0$.
This implies that the GP must have an exponent of $\eta =1$. Indeed, for a
GP satisfying Eq.~(\ref{GP})
we have $\!\!\!\overline{[h(x_{n+2})-h(x_{n+1})][h(x_{n+1})-h(x_{n})]}%
\propto (2^{\eta -1}-1)(\delta x)^{\eta }.$ That correlator vanishes if and
only if $\eta =1$. 

The solid lines in Fig.~\ref{scaled} show that our approximation works
even though we do not deal with a true GP ($w^{(n)}$ carries zeros in row
and column $(n+1)$). To see why, we omit on the right--hand side of Eq.~(\ref%
{HFK}) in $h^{(n)}$ the eigenvalue $\epsilon _{n+1}^{(n)}$ and in $%
w^{(n)}$ row and column labelled $n+1$ and perform the step $%
n\rightarrow (n+1)$. This procedure is repeated with $n$ replaced by $(n+1)$%
, etc. We obtain a genuine GP except for the spacing of the eigenvalues $%
\epsilon _{\alpha }^{(n)}$. These obey Wigner--Dyson statistics for $\alpha
\leq n$ and for $\alpha \geq (n+1)$ (see Ref.~\cite{alhassid05}) while the
HF gap separating these two sets of levels does not. Because of the gap the
process is not a GP and we expect less mixing between levels below and above
the gap than would otherwise occur.
We thus expect (on top of the step seen in the data) a slowing--down of the
decorrelation in the vicinity of the Fermi level. It appears that this
effect is too small to rise above the statistical uncertainties.

\emph{Scaling of $n$.} Universal correlations are obtained by scaling the
parameter $x$. The scaled parameter for a GP with exponent $\eta$ is
generally given by $x_{\mathrm{eff}}=D^{1/\eta }x$ where $D=\left[ \overline{%
(\tilde{\epsilon}_{\alpha }(x+\delta x)-\tilde{\epsilon}_{\alpha }(x))^{2}}%
/\delta x^{\eta }\right] $. We identify $x$ with $n$ and use that relation
for the smallest possible change $\delta n=1$ to define (for $\eta=1$) the scaled parameter
$n_{\mathrm{eff}}=\overline{{(\tilde{\epsilon}_{\alpha }^{(n+1)}-\tilde{\epsilon}%
_{\alpha }^{(n)})}^{2}}.$ Assuming that $\Delta (\epsilon _{\alpha })$
does not change much when one electron is added and using Koopmans' limit,
we estimate $\tilde{\epsilon}_{\alpha }^{(n+1)}-\tilde{\epsilon}_{\alpha
}^{(n)}\approx (\epsilon _{\alpha }^{(n+1)}-\epsilon _{\alpha
}^{(n)})/\Delta (\epsilon _{\alpha })\approx v_{\alpha n+1;\alpha
n+1}^{A}/\Delta (\epsilon _{\alpha })$. Since $\sigma ^{2}(v_{\alpha
n+1;\alpha n+1}^{A})=4u^{2}$, we obtain the simple approximate scaling given
in Eq.~(\ref{deltan}).
In general, the mean level spacing in Eq.~(\ref{deltan}) depends (weakly) on
$n$. Although $n$ is a discrete parameter, the scaled parameter $n_{\mathrm{%
eff}}$ can be made as close to being continuous as we like by choosing a
sufficiently small $u$.

\begin{figure}[h]
\centerline{\includegraphics[scale=0.9]{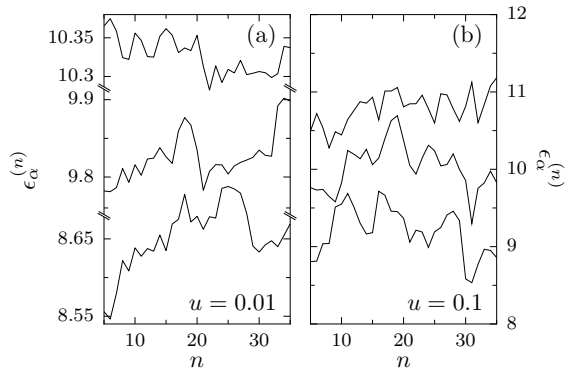}}
\caption{The addition of electrons to a quantum dot is analogous to Brownian
motion. Shown are three HF single--particle
levels $\epsilon _{\alpha}^{(n)} $ versus the number $n$
of electrons for $u=0.1$ (right) and $u=0.01$ (left). Notice the finer scale
on the left. }
\label{scramble}
\end{figure}
\emph{Discussion.} The solid lines in the top (bottom) panels of Fig.~\ref%
{scaled} display the two universal GP curves for $o(\delta x)=\overline{%
|\langle \psi (x+\delta x)|\psi (x)\rangle |^{2}}$ and for $\sigma (\alpha
,\delta x)$. These are computed numerically from an $\eta =1$ GP whose
parameter $x$ is scaled according to the prescription given in the previous
paragraph and identified with $\delta n_{\mathrm{eff}}$. The wave--function
correlator can be well fit by $o(\delta n_{\mathrm{eff}})\approx \lbrack
1+\delta n_{\mathrm{eff}}/\alpha ]^{-1}$ with $\alpha =0.23$~\cite{rmp00}.
The steps in the data for $\alpha =14\ (18)$ at $\delta n=4\ (8)$ can be
understood in the framework of the generalized Koopmans approximation~(\ref%
{HFK}). In the HF approximation of the $n$--electron dot, the level $\alpha
=(n+1)$ is empty.
In the approximation~(\ref{HFK}) this level is not coupled by the
perturbation $w^{(n)}$ to any other HF eigenstate $\gamma $ (we recall that $%
w_{\gamma ,n+1}^{(n)}=w_{n+1,\gamma }^{(n)}=0$) and is, therefore, also an
eigenstate of $h^{(n+1)}$. In particular, $\epsilon _{\alpha
=n+1}^{(n+1)}\approx \epsilon _{\alpha =n+1}^{(n)}$ and $\psi _{\alpha
=n+1}^{(n)}\approx \psi _{\alpha =n+1}^{(n+1)}$. This is why the measures
for both spectral and wave--function scrambling remain approximately
unchanged between $n=\alpha -1$ and $n=\alpha $.

When plotted versus $n$, the single--particle HF levels should exhibit
Brownian--motion--like behavior since the addition of electrons is an $\eta
=1$ GP
even for very small $u$. This is verified in Fig.~\ref{scramble} where we
show three HF levels versus $n$ for a particular realization of the induced
ensemble. When $u$ is reduced from $u=0.1$ (right) to $u=0.01$ (left), the
behavior remains non--analytic although on a much finer scale.

In conclusion, we have shown that, in a chaotic or diffusive dot with
interactions, the scrambling of HF levels due to the addition of electrons
is described by a Brownian--motion type process. The change of the number $n$
of electrons is analogous to a discrete equally--spaced time series. We have
used a non--universal scaling of $n\rightarrow n_{\mathrm{eff}}$ to show
that the measures of scrambling of both HF levels and HF wave functions
versus $n_{\mathrm{eff}}$ follow closely the corresponding universal
correlators of an $\eta =1$ Gaussian random process. The only exception
occurs when an empty level becomes filled, causing a delay in the scrambling
process by one electron.

 This work was supported in part by the U.S. DOE grant No.\
DE-FG-0291-ER-40608. We thank Y. Gefen and Ph. Jacquod for 
useful discussions. Y.A. acknowledges support by a von Humboldt Senior
Scientist Award and the hospitality of the Max-Planck-Institut f\"ur
Kernphysik at Heidelberg where part of this work was completed.

\end{document}